\documentclass[prd,aps,eqsecnum,amsmath,floatfix,nofootinbib,preprint,tightenlines]{revtex4}

\usepackage{latexsym}
\usepackage{graphicx}
\usepackage{multirow}
\usepackage[dvipsnames]{xcolor}

\usepackage{hyperref}

\def\ttl#1{{\it #1}}

\def\NON{\nonumber\\}
\def\bibi{\bibitem}

\def\a{\alpha}

\def\c{\chi}
\def\d{\delta}
\def\f{\phi}                    
\def\g{\gamma}
\def\h{\eta}

\def\j{\psi}

\def\l{\lambda}
\def\m{\mu}
\def\n{\nu}

\def\p{\pi}                     
\def\th{\theta}                  
\def\r{\rho}                    
\def\s{\sigma}                  

\def\J{\Psi}

\def\O{\Omega}
\def\P{\Pi}

\def\S{\Sigma}

\def\ca{{\cal A}}

\def\cd{{\cal D}}

\def\cl{{\cal L}}
\def\cm{{\cal M}}

\def\cu{{\cal U}}
\def\cv{{\cal V}}

\def\cbo{{\,\raise-.15ex\Sc [\,}}                       

\def\Sl#1{\rlap{\hbox{$\mskip 3 mu /$}}#1}      

\def\svev#1{\left\langle #1\right\rangle}       

\def\ddt#1{{\buildrel {\hbox{\LARGE .\kern-2pt.}} \over {#1}}}

\def\ie{\mbox{\it i.e.}}
\def\eg{\mbox{\it e.g.}}

\def\tr{{\rm tr}\,}
\def\Tr{{\rm Tr}\,}
\def\hc{{\rm h.c.\,}}

\def\half{{1\over 2}}

\def\Im{{\rm Im\,}}

\def\det{{\rm det}}

\def\seef{{\it cf.\  }}

\def\tr{\,{\rm tr}}

\def\bj{\overline{\j}}
\def\bs{\overline{\s}}

\def\bJ{\overline{\J}}
\def\ta{\tilde{\a}}
\def\tO{\tilde\O}

\def\tm{\tilde\m}
\def\hp{\hat\p}
\def\tilt{\tilde\theta}
\def\tj{{\tilde{j}}}

\def\tU{{\tilde{U}}}

\def\id{{\bf 1}}

\def\irrep{{\it irrep}}
\def\irreps{{\it irreps}}

\def\Nf{N}

\def\theff{\th_{\rm eff}}
\def\thshft{\th_{\rm shf}}
\def\thind{\th_{\rm ind}}
\def\Nmaj{N_{\rm maj}}

\def\Dslash{\Sl{D}}
\def\SU{\textrm{SU}}
\def\SO{\textrm{SO}}
\def\Uone{\textrm{U(1)}}

\def\pf{\textrm{pf}}

\begin{document}

\begin{center}
{\large\bf Phase ambiguity of the measure for \\[2mm]
  continuum Majorana fermions}\\[8mm]
Maarten Golterman$^{a,b}$ and Yigal Shamir$^c$\\[10 mm]
{\small
$^a$Department of Physics and IFAE-BIST, Universitat Aut\`onoma de Barcelona\\
E-08193 Bellaterra, Barcelona, Spain\\
$^b$Department of Physics and Astronomy, San Francisco State University,\\
San Francisco, CA 94132, USA\\
$^c$Raymond and Beverly Sackler School of Physics and Astronomy,\\
Tel~Aviv University, 69978, Tel~Aviv, Israel}\\[10mm]
\end{center}

\begin{quotation}
Integrating over a continuum Majorana fermion formally yields
a functional pfaffian.  We show that the phase of this pfaffian
is ambiguous, as it depends on the choice of basis.
This ambiguity is naturally resolved within a non-perturbative lattice definition,
allowing us to discuss the relation between the phase of the lattice pfaffian
and the effective $\th$ angle of the theory.  We also resolve an
apparent paradox regarding the induced $\th$ angle when a theory
of $N$ Dirac fermions in a real representation of the gauge group
is re-expressed in terms of $2N$ Majorana fermions.
We discuss how all this is reflected in chiral perturbation theory.
\end{quotation}

\newpage
\section{\label{intro} Introduction}
In a QCD-like theory with Dirac fermions, the measure of
the euclidean functional integral is positive when all fermions
have a positive mass, and, as a consequence, there is no topological term
induced by the fermionic part of the theory.
This generalizes to all types of fermion \irreps:
complex, real, and pseudoreal.

If the theory contains $N$ Dirac fermions in a real \irrep,
we may reformulate it in terms of $2N$ Majorana fermions.
We will be using Majorana fields each of which
packs together a Weyl fermion and its anti-fermion.\footnote{
  The precise definition is given in Eq.~(\ref{Diracreal}) below.}
Assuming an equal positive mass $m>0$ for all Dirac flavors,
the mass matrix $M$ of the Majorana formulation
is then given by $M=mJ_S$,
with the $2N \times 2N$ matrix
\begin{equation}
  J_S = \left( \begin{array}{cc}
    0 & \id_N \\ \id_N & 0
  \end{array} \right)\ ,
\label{Js}
\end{equation}
where $\id_n$ is the $n\times n$ unit matrix.
A non-anomalous chiral rotation can then be used to bring the mass matrix
to a flavor-diagonal form $M=m J_S^{\rm rot}$ where
\begin{equation}
  J_S^{\rm rot} = i\g_5 \id_{2N} \ ,
  \label{Js2}
\end{equation}
showing that each entry of $M$ has a $\Uone_A$ phase equal to $\p/2$.
Now let us apply a $\Uone_A$ rotation that turns
the mass matrix into a positive matrix, $M=m\id_{2N}$.  Because of the anomaly,
this generates a topological term $e^{i\th Q}$, where
\begin{equation}
\label{topo}
  Q = \frac{g^2}{32\p^2}\int d^4x \tr(F\tilde{F})\ ,
\end{equation}
is the topological charge, and
\begin{equation}
\label{pihalf}
  \th = -\p N T/2 \ ,
\end{equation}
with $T$ the index of the Dirac operator for the fermion \irrep\
in a single instanton background.

Let us consider the consequences of this topological term.
$T$ is always even for a real \irrep.\footnote{%
   We will recover this result in Sec.~\ref{maj}.}
If $NT$ is divisible by 4
then $e^{i\th Q}=1$, and the topological term drops out.
If $NT$ is not divisible by 4, we have $e^{i\th Q}=(-1)^Q$.
Hence, it appears that the Majorana measure will be positive for $Q$ even,
but negative for  $Q$ odd.  This is puzzling,
because the measure of the original Dirac theory is positive for any $Q$,
and, obviously, the Dirac and Majorana formulations should represent
the same theory.

The paradox would be resolved if the very transition to the
Majorana formulation would somehow generate a ``compensating'' topological term
$e^{i\p NT Q/2}$.  The additional topological term induced by
the $\Uone_A$ rotation would then cancel against the
compensating topological term.  We would end up with a positive mass matrix
and with no topological term, as in the original Dirac theory.

The purpose of this paper is to show that this is indeed what happens.
In reality, it turns out that the paradox described above arises
because in the argument we ignored a phase ambiguity
of the Majorana measure which is present in the formal continuum theory.
The existence of this ambiguity allows us to {\em require} agreement between
the Dirac and Majorana formulations.  When the Majorana mass matrix
involves $J_S$ or $J_S^{\rm rot}$, this requirement implies the existence of
the compensating topological term in the path integral.
Going beyond formal arguments, we demonstrate the presence of
the compensating topological term through a fully non-perturbative
lattice derivation of the transition from the Dirac to the Majorana
formulation.  Finally, we discuss the implications
for the chiral effective theory.

This paper is organized as follows.   In Sec.~\ref{maj} we show how, in the continuum, a phase
ambiguity arises in the choice of a basis for a gauge theory with Majorana fermions.
We explain how this ambiguity can be resolved
in a theory with an even number of Majorana fermions
by comparison with the same theory formulated in terms of Dirac fermions.
Then, in Sec.~\ref{latt}, we show that the lattice formulation implies a natural choice of
basis, thus fixing the phase consistently, both in the formulations with Wilson and with
domain-wall fermions.   This allows us to discuss the $\th$ angle induced by the lattice
fermion action, reviewing and generalizing the earlier work of Ref.~\cite{SSt}.
We consider separately a stand-alone gauge theory of Majorana fermions,
and a theory of $2N$ Majorana fermions obtained by reformulating
a theory of $N$ Dirac fermions.
We then revisit the precise form of the condensate in the presence of a fermion-induced
$\th$ angle, both
in the gauge theory as well as in chiral perturbation theory.   This is done in
Sec.~\ref{dirac} for a theory with Dirac fermions in a complex \irrep\ of the gauge
group, and in Sec.~\ref{vacreal} for a theory with Majorana fermions in a real \irrep\
of the gauge group.   Section~\ref{conc} contains our summary and conclusion.
There are six appendices dealing with technical details.

\section{\label{maj} Majorana fermions and the phase ambiguity}
In this section, we first review some useful standard results for Dirac
(Sec.~\ref{diracrev}) and Majorana (Sec.~\ref{majrev}) fermions.
We then discuss the phase ambiguity that is encountered
in defining the continuum path integral for Majorana fermions
(Sec.~\ref{ambg}).

\subsection{\label{diracrev} Dirac fermions}
Consider a euclidean gauge theory with $N$ Dirac fermions in some \irrep\ of
the gauge group.  The partition function for the most general
choice of parameters is
\begin{equation}
  Z = \int \cd A \cd\j\cd\bj\, \exp\left(-\int d^4x\, \cl \right)\ ,
\label{Z}
\end{equation}
where
\begin{equation}
\label{lag}
  \cl = \frac{1}{4} F^2 + \bj (\Sl{D} + \cm^\dagger P_L + \cm P_R ) \j
  + i \th Q \ ,
\end{equation}
with $P_{R,L}=(1\pm\g_5)/2$, and $\cm$ is a complex $N\times N$ matrix.
The topological charge $Q$ was introduced
in Eq.~(\ref{topo}).  We will specialize to a mass matrix of the form
\begin{equation}
\label{cm}
  \cm = m\O = m e^{i\a/(NT)} \tO \ , \qquad \tO \in \SU(\Nf) \ ,
\end{equation}
with real $m>0$ and a real phase $\a$.  Upon integrating out the fermions the dependence on $\tO$
drops out thanks to the invariance under non-singlet chiral transformations,
and
\begin{equation}
\label{DetF}
  \det(\Sl{D} + \cm^\dagger P_L + \cm P_R)
  = e^{i\a Q}\, \det^N(\Sl{D}+m) \ ,
\end{equation}
where, on the right-hand side, $\Sl{D}+m$ is the one-flavor Dirac operator.
This result can be derived using the spectral representation
of the Dirac operator, see App.~\ref{specrep}.
As mentioned earlier, $T$ is the index of the Dirac operator
in a single instanton background.
The measure $\m(A)$ of the path integral is thus
\begin{eqnarray}
\label{msr}
  \m(A) &=& e^{-i\theff Q}\, e^{-\frac{1}{4} F^2} \det^N(\Sl{D}+m)
\\
  &=& e^{-i\theff Q}\, \tm(A) \ ,\nonumber
\end{eqnarray}
where
\begin{equation}
\label{mutilde}
  \tm(A) = e^{-\frac{1}{4} F^2} \det^N(\Sl{D}+m) \ ,
\end{equation}
is positive, and the effective topological angle is
\begin{equation}
\label{theff}
  \theff = \th - \a \ .
\end{equation}

\subsection{\label{majrev} Majorana fermions}
A theory of $N$ Dirac fermions
in a real representation of the gauge group $G$ can be reformulated in terms of
$2N$ Majorana fermions.
The $N$ Dirac fermions are composed of $2N$ Weyl fermions.
From these Weyl fermions, we construct Majorana fermions each of which
packs together a Weyl fermion and its
anti-fermion, which is possible because the fermion and
the anti-fermion belong to the same representation of $G$.

The mapping between Dirac fermions (on the right-hand side)
and Majorana fermions (on the left-hand side) is
\begin{eqnarray}
\label{Diracreal}
  \J_{L,i}    &=&   \j_{L,i}      \ ,\\
  \J_{R,i}   &=&   CS\bj^T_{L,i}     \ ,\nonumber\\
  \J_{R,N+i}  &=&   \j_{R,i}      \ ,\nonumber\\
  \J_{L,N+i} &=&   CS\bj^T_{R,i}    \ ,\nonumber
\end{eqnarray}
where $i=1,\ldots,N$.  Here
$C$ the charge conjugation matrix, and $S$
the group tensor satisfying the invariance property $g^TSg=S$ for all $g\in G$.
We recall the basic properties, $C^{-1}=C^\dagger=C^T=-C$,
and $S^{-1}=S^\dagger=S^T=S$.  We also introduce
\begin{equation}
\label{majcond}
  \bJ \equiv \J^T CS \ .
\end{equation}
Thus, Eq.~(\ref{Diracreal}) determines all the components of the Majorana fermions
in terms of the original Dirac fermions, or, equivalently,
in terms of the corresponding Weyl fields.
Other mappings between Dirac and Majorana fermions are
possible, and we give an example in App.~\ref{bases}.
What is special about Eq.~(\ref{Diracreal}) is that it respects
the natural mapping between Weyl and Majorana fields.

Proceeding to the lagrangian, for the kinetic term we have
\begin{equation}
\label{LK}
  \cl_K = \sum_{i=1}^N \bj \Dslash \j = \half\sum_{I=1}^{2N} \bJ_I\Dslash\J_I \ .
\end{equation}
For the mass term we have
\begin{equation}
\label{LM}
  \cl_m = m \bj e^{i\a_D\g_5} \j = \frac{m}{2}\, \bJ e^{i\a_D\g_5} J_S \J \ ,
\end{equation}
where the $2N\times 2N$ matrix $J_S$ was introduced in Eq.~(\ref{Js}),
and $\a_D=\a/(NT)$ is the phase introduced in the Dirac case in Eq.~(\ref{cm}).
We have set $\tO=1$, since the $\SU(N)$ part of the original Dirac mass matrix
does not play a role in the following.

The flavor symmetry is as follows.  In the massless limit,
the theory is invariant under $\SU(2N)$ transformations
\begin{eqnarray}
\label{transfreal}
  \J  &\to& \left(P_L h + P_R h^*\right)\J\ ,\\
  \bJ &\to& \bJ (P_L h^T + P_R h^\dagger)\ ,\nonumber
\end{eqnarray}
with $h\in\SU(2N)$.  When the mass term~(\ref{LM}) is turned on,
the $\SU(2N)$ symmetry is explicitly broken to $\SO(2N)$.
The Dirac formulation of the same theory obviously has
the same global symmetry; but the full symmetry is manifest
only in the Majorana formulation.\footnote{%
  For a discussion of how the global symmetry is realized in
  the Dirac formulation, see Ref.~\cite{sextet}.
}

\subsection{\label{ambg} Pfaffian phase ambiguity}
There exists a non-anomalous
$\SU(2N)$ chiral rotation that brings the Majorana mass term~(\ref{LM})
to a diagonal form
\begin{equation}
\label{massDiag}
  \cl_m = \frac{m}{2}\, \bJ i\g_5 e^{i\a_D\g_5} \J
        = \frac{m}{2}\, \bJ e^{i(\a_D+\p/2)\g_5} \J \ .
\end{equation}
We see that we have an extra $\Uone$ phase of $\p/2$, leading to an apparent
paradox, as explained in the introduction.   In the following,
we ask the question of how this paradox may be
resolved in the continuum.  In Sec.~\ref{latt} we will show
how it is avoided, by introducing a non-perturbative regulator.

To start, let us consider a single Majorana fermion
with lagrangian
\begin{eqnarray}
\label{L1}
  \cl &=& \half \bJ D \J \ = \ \half \J^T CSD\, \J\ ,
\\
\label{Dm}
  D &=& \Dslash + m e^{i\a_M\g_5} \ .
\end{eqnarray}
The differential operator $CSD$ is antisymmetric, and the result of
formally integrating out the Majorana fermion is $\pf(CSD)$,
the pfaffian of $CSD$.

In the Dirac case,
$\det(D)$ is simply equal to the (regulated) product of all eigenvalues,
see App.~\ref{specrep}.  What about pfaffians?

Introducing the abbreviation $\ca=CSD$, the effect of a unitary
change of basis for Majorana fermions is
\begin{equation}
\label{AA'}
  \ca \to \ca' = \cu^T \ca\, \cu \ ,
\end{equation}
where both $\ca$ and thus $\ca'$ are antisymmetric.
We will be looking for a change of basis so that $\ca'$ will have
a skew-diagonal form.

For a real representation,
the eigenvalues of the Dirac operator have a twofold degeneracy.
Because its hermitian part is equal to $m\cos\a_M$ times the identity matrix,
the Dirac operator~(\ref{Dm}) is normal, $[D,D^\dagger]=0$.
Consider an eigenvector $\c$ with eigenvalue $\l$.
By normality, $D\c=\l\c$ implies $D^\dagger\c=\l^*\c$.  Hence
\begin{equation}
  D\,CS\c^* = CS D^T \c^* = CS (D^\dagger\c)^* = CS(\l^*\c)^* = \l CS\c^*\ .
\label{CSpsi}
\end{equation}
It follows that $CS\c^*$ is an eigenmode with the same eigenvalue as $\c$.
The eigenmodes $\c$ and $CS\c^*$ are orthogonal,
$(CS\c^*)^\dagger \c = -\c^T CS \c =0$,
where we used that the matrix $CS$ 
is antisymmetric.

The skew-diagonal representation $\ca'$ is achieved by transforming
to a basis in which each eigenvector $\c$
is followed by its companion eigenvector $CS\c^*$.
Selecting arbitrarily one eigenvector from each pair,
and labeling the resulting subset as $\c_1,\c_2,\ldots$,
we consider the unitary change of basis generated by the matrix $\cu$
whose columns are comprised of the ordered pairs of eigenvectors,
\begin{equation}
\label{Ubasis}
  \cu = (\c_1, e^{i\f_1}CS\c_1^*, \c_2, e^{i\f_2}CS\c_2^*, \ldots) \ .
\end{equation}
Notice that, for each pair, we have allowed the second eigenvector
to have an arbitrary U(1) phase relative to the original eigenvector.
These arbitrary phases play a profound role, as we will now see.

The $2\times2$ subspace of $\ca'$ associated with a pair $\c, e^{i\f}CS\c^*$
with eigenvalue $\l$ has the explicit form
\begin{equation}
  \left(\begin{array}{c} \c^T \\ e^{i\f}\c^\dagger SC^T \end{array} \right)
  CS D
  \left(\begin{array}{cc} \c & e^{i\f}CS\c^* \end{array} \right)
  = e^{i\f} \l \left(\begin{array}{cc} 0 & -1 \\ 1 & 0 \end{array} \right) \ .
\label{2by2phi}
\end{equation}
The pfaffian of $\ca'$ factorizes as the product of pfaffians
for the $2\times2$ subspaces, where the pfaffian of the
above $2\times2$ subspace is, by definition, equal to $-e^{i\f} \l$.    Explicitly,
\begin{equation}
\label{pfA'}
  \pf(\ca') = \prod_i (-e^{i\f_i} \l_i) \ .
\end{equation}
This result exhibits a phase ambiguity, represented by the sum
$\sum_i\f_i$.

In retrospect, the phase ambiguity can be traced to the elementary
property $\pf(\ca') = \pf(\ca) \det(\cu)$.
This relation implies that the phase of the pfaffian depends on the
choice of basis for the Majorana field on which the differential operator
$\ca$ acts.  The basis is represented by the unitary matrix $\cu$,
and $\det(\cu)$ is, thus, a basis-dependent phase.

The rigorous resolution of the phase ambiguity requires
a non-perturbative treatment in order to specify the basis, which we will give in Sec.~\ref{latt}.
In the rest of this section we restrict ourselves to an
even number of Majorana fermions, and discuss how the phase
may be fixed by appealing to the corresponding theory
defined in terms of Dirac fermions, where no such phase ambiguity exists.

As reviewed in App.~\ref{specrep} for the Dirac case, let us consider separately the zero modes
and the non-zero modes.  Starting with the non-zero modes, and following App.~\ref{specrep}, the eigenvectors
$\c_\pm$ now each have a companion, $e^{i\f_\pm}CS\c_\pm^*$,
where we have allowed for arbitrary relative U(1) phases.
The contribution of these two pairs of eigenvectors to $\pf(\ca)$ is
\begin{equation}
\label{twopairs}
  (-e^{i\f_+} \l_+)(-e^{i\f_-} \l_-) = e^{i(\f_++\f_-)} (\l^2+m^2) \ ,
\end{equation}
where, as in App.~\ref{specrep}, $\l^2$ is the eigenvalue
of the second-order operator $-\Dslash^2 P_R$.
For a single Dirac fermion in the same real representation,
the contribution of the eigenvectors $\c_\pm$ and $e^{i\f_\pm}CS\c_\pm^*$
to $\det(D)$ is simply a factor of
\begin{equation}
\label{detnonzero}
  (\l^2+m^2)^2 \ .
\end{equation}
The determinant is independent of the arbitrary U(1) phase of each eigenvector.
If we now take two Majorana fermions,
the corresponding contribution to $\pf(\ca')$ will be
\begin{equation}
\label{pfnonzero}
  [e^{i(\f_++\f_-)} (\l^2+m^2)]^2 \ .
\end{equation}
We see that, by making the {\em choice}
\begin{equation}
\label{phasenonzero}
  \f_+ = \f_- = 0 \ ,
\end{equation}
we achieve agreement between the corresponding factors
for the Dirac and two-Majorana cases.\footnote{It is, in fact, sufficient to choose
$\f_++\f_-=0\ \mbox{mod}\ \p$.}

Proceeding to the zero modes, in the Dirac case the contribution
of a pair of zero modes, $\c_0,e^{i\f_0}CS\c_0^*$, is just
\begin{equation}
\label{detzero}
  (m e^{\pm i\a_D})^2 \ ,
\end{equation}
depending on the chirality.  In the Majorana case,
the corresponding contribution to $\pf(\ca')$ from each Majorana fermion is
\begin{equation}
\label{onepfzero}
  -m e^{i\f_0} e^{\pm i\a_M} =  -m e^{i\f_0} e^{\pm i(\a_D+\p/2)} \ ,
\end{equation}
where on the right-hand side we have substituted $\a_M=\a_D+\p/2$.
The extra phase of $\p/2$ arises during the transition from the Dirac
to the Majorana formulation, as we have seen
in Eq.~(\ref{massDiag}).  The contribution from two Majorana fermions is thus
\begin{equation}
\label{twopfzero}
  (m e^{i\f_0} e^{\pm i(\a_D+\p/2)})^2 \ .
\end{equation}
It follows that the Dirac result~(\ref{detzero}) will only be reproduced provided we make
the non-trivial choice
\begin{equation}
\label{phasezero}
  \f_0 = \p/2\ \mbox{mod}\ \p \ .
\end{equation}

\section{\label{latt} Non-perturbative calculation}
In the previous section, we showed that the definition of a theory with
Majorana fermions has an intrinsic phase ambiguity, which can be used
to resolve the apparent paradox introduced in Sec.~\ref{intro}.
However, the question of whether, and how, the theory ``chooses''
the proper phase was left open.
In order to address this question, we need a properly regulated
non-perturbative definition of the theory, which is provided by the lattice.

The lattice action for a Majorana fermion will always have the generic
form $\half\J^T \ca \J$ for a suitable antisymmetric operator $\ca$.
Integrating over the lattice Majorana field yields $\pf(\ca)$,
which is now well defined.  There is no room for any (phase) ambiguity,
because, on any finite-volume lattice, $\ca$ is a finite-size matrix,
and the lattice selects the coordinate basis to define $\ca$.

Our first result concerns a single Majorana fermion
with no chiral angle(s), and a positive bare mass $m_0>0$.
Using domain-wall fermions, we show in App.~\ref{pospf} that $\pf(\ca)$ is
strictly positive in this case.  The domain-wall fermion measure
is then strictly positive for any number of Majorana fermions,
and in all topological sectors.

In this section, we discuss in detail the transition from the Dirac
to the Majorana formulation.
In Sec.~\ref{Wilson}, we regulate the theory using Wilson fermions, and
in Sec.~\ref{DWF}, using domain-wall fermions.
 While in the case of Wilson
fermions, there is a lacuna in the argument (which we discuss in some detail in
App.~\ref{SStconj}), this is not the
case for domain-wall fermions.
We establish that the compensating topological term alluded to
in the introduction indeed arises when needed, thus resolving the paradox.
As in App.~\ref{pospf},
it proves easier to work with the five-dimensional formulation
of domain-wall fermions, rather than directly with any Ginsparg-Wilson operator
that arises in the limit of an infinite fifth dimension.
We also remark that staggered
fermions always lead to a four-fold taste degeneracy in the continuum
limit, and so they cannot be used here, given that the apparent paradox
only arises when $NT$ is even, but not divisible by four.\footnote{
  Interpreting the staggered tastes as physical flavors,
  it is possible that reduced staggered fermions can be
  employed \cite{STW1981,rdcstag}.   We have not explored this further.}
We summarize the results of this section in Sec.~\ref{smr}.

\subsection{\label{Wilson} Wilson fermions}
If we formulate the theory using Wilson fermions, the resolution of the puzzle relies on the observation, made in
Ref.~\cite{SSt}, of how the $\th$ angle can be realized within this fermion
formulation.  The starting point of the discussion is a one-flavor
Wilson operator with both the Wilson and mass terms chirally rotated by angles $\th_W$
and $\th_m$, respectively,
\begin{equation}
  D_W(\th_W,\th_m) = D_K + e^{i\th_W\g_5} W + e^{i\th_m\g_5} m_0 \ .
\label{Wth}
\end{equation}
Here $D_K$ is the naive lattice discretization of the (massless)
Dirac operator.
$W$ is the Wilson term, which eliminates the fermion doublers,
and is chosen for definiteness to be
real positive; $m_0$ is the bare mass.  The partition
function takes the form~(\ref{Z}), but with the fermion part of the lagrangian replaced
by\footnote{We will not need the lattice form of the gauge action.}
\begin{equation}
  \cl_F = \bj D_W(\th_W,\th_m) \j \ .
\label{LFth}
\end{equation}
First, only the difference $\th_W-\th_m$ can
be physical, as can be seen by applying the transformation
\begin{equation}
  \j\to e^{i\h\g_5} \j \ , \qquad \bj\to \bj e^{i\h\g_5} \ .
\label{rot5}
\end{equation}
In the lattice regulated theory, the determinant
of this transformation is unity, hence it provides an alternative
representation of exactly the same theory.  It is easily checked
that this transformation leaves the $D_K$ part invariant,
while the angles undergo the transformation $\th_W\to\th_W+2\h$,
$\th_m\to\th_m+2\h$.  By choosing $\h=-\th_m/2$
we eliminate the phase of the mass term, while the phase
of the Wilson term becomes $\th_F\equiv\th_W-\th_m$.

With only the angle $\th_F$ left in the fermion action,
and with $\th$ as the explicit vacuum angle (see Eq.~(\ref{lag})), what
Ref.~\cite{SSt} claimed is that, in the continuum limit,
\begin{equation}
  Z(\th,\th_F) = Z(\th+NT\th_F,0) \ .
  \label{Zlim}
\end{equation}
This implies that the relative $\Uone$ phase of the Wilson term
and the mass term turns into the familiar $\th$ angle in the continuum limit.
In Eq.~(\ref{Zlim}) we have written down the generalization
of the result of Ref.~\cite{SSt} to $N$ Dirac fermions in an \irrep\ with
index $T$. In the case that a topological term with $\th\ne 0$
is already present in the gauge action, $NT\th_F$ gets added to $\th$.

We pause here to note that the argument given in Ref.~\cite{SSt}
is not complete as it stands, because of a subtlety related to renormalization.
While it is beyond the scope of this paper to complete the
proof, App.~\ref{SStconj} outlines a conjecture on the interplay of the
observation of Ref.~\cite{SSt} and renormalization.   However,
this subtlety does not affect the rest of this paper.   In particular, in the
next subsection we provide an argument analogous to the one given here
based on domain-wall fermions, where the subtlety does not arise.

Next, let us work out the transition from the Dirac to the Majorana case.
We start with a single Dirac fermion in a real \irrep, where the Wilson
fermion operator $D_W(\th_F)$ is given by Eq.~(\ref{Wth}), taking $\th_W=\th_F$ and
$\th_m=0$.  In the Majorana formulation,
the $4\times 4$ matrix in spinor space becomes an $8\times 8$ matrix
which mixes the two Majorana species.
In terms of $4\times 4$ blocks, the Wilson operator in the Majorana formulation
is
\begin{equation}
  D_{\rm Maj}(\th_F) = \left( \begin{array}{cc}
    D_K & e^{i\th_F\g_5} W + m_0 \\
    e^{i\th_F\g_5} W + m_0 & D_K
  \end{array} \right) \ ,
\label{Dmaj1}
\end{equation}
where we have used Eqs.~(\ref{Diracreal}) and~(\ref{majcond}).
The lagrangian becomes
\begin{equation}
  \cl_F = \half \bJ D_{\rm Maj}(\th_F) \J \ .
\label{Lmaj}
\end{equation}

The key feature of Eq.~(\ref{Dmaj1}) is that,
because of their identical chiral properties,
the Wilson and mass terms occur in the same places.  Applying an $\SU(2)$
flavor transformation, \ie, using Eq.~(\ref{transfreal}) for $N=1$ with $h=\exp(-i\p\s_2/4)=h^*$,
and using that $h^T\s_1 h = \s_3$,
the Majorana Wilson operator gets rotated into
\begin{equation}
  D_{\rm Maj}(\th_F) = \left( \begin{array}{cc}
    D_K + e^{i\th_F\g_5} W + m_0 & 0 \\
    0 & D_K - (e^{i\th_F\g_5} W + m_0)
  \end{array} \right) \ .
\label{Dmaj3}
\end{equation}
When $\th_F=0$, the relative phase of the Wilson and mass
terms is zero, for both of the Majorana species.  This implies
that $D_{\rm Maj}(0)$ is the Wilson operator
for two Majorana fermions with the same bare mass $m_0$
(as opposed to the case where one Majorana fermion would have a mass $+m_0$
and the other $-m_0$).

We prove this assertion by applying the transformation~(\ref{rot5})
with $\h=\p/2$ to the second Majorana fermion only.\footnote{
  Note that the transformation~(\ref{rot5}) is consistent
  with the Majorana condition~(\ref{majcond}).
}
Explicitly, it reads $\J_2 \to i\g_5\J_2$.
The Majorana--Wilson operator transforms into
\begin{eqnarray}
  D_{\rm Maj}(\th_F) &\to& \left( \begin{array}{cc}
    D_K + e^{i\th_F\g_5} W + m_0 & 0 \\
    0 & i\g_5 \Big( D_K - (e^{i\th_F\g_5} W + m_0) \Big) i\g_5
  \end{array} \right)
\label{finalWmaj}\\
  &=& \rule{0ex}{5ex} \left( \begin{array}{cc}
    D_K + e^{i\th_F\g_5} W + m_0 & 0 \\
    0 & D_K + e^{i\th_F\g_5} W + m_0
  \end{array} \right) \ .
\nonumber
\end{eqnarray}
The fermion operator for each Majorana fermion
is now exactly the same as in the Dirac case.
The corresponding basis for the Majorana fields is given in App.~\ref{bases}.
It follows that the fermion measure of the two-Majorana formulation is equal to  $\pf^2(CSD_W(\th_F))$, and thus equal to the
Dirac measure $\det(D_W(\th_F))$.  We have proved
that the fermion measure in the Majorana formulation is unchanged from the Dirac
formulation.

Equation~(\ref{finalWmaj}) shows that we can choose the mass matrix
to be proportional to the unit matrix, instead of to $J_S$ (Eq.~(\ref{Js}))
or $J_S^{\rm rot}$ (Eq.~(\ref{Js2})).
Unlike in the formal continuum treatment of the previous section,
no phase ambiguity, nor any ``excess'' phase of $\p/2$, arises
when the transition to Majorana fermions is done
in the lattice-regulated theory.

\subsection{\label{DWF} Domain-wall fermions}
In this subsection, we revisit the argument
of the previous subsection,  but now using domain-wall fermions \cite{DBK} instead of Wilson fermions.
As we will see, in the case of domain-wall fermions, the argument is complete, allowing us to
conclude that a lattice regularization can indeed be invoked to settle the ambiguity we found in
Sec.~\ref{maj}.

The starting point is the domain-wall fermion action \cite{YSdwf1} for a massive Dirac fermion with bare mass $m_0$ and domain-wall height $M$,
\begin{eqnarray}
\label{Sdwf}
S&=&\sum_{s=1}^{N_5} \bj(s)(D_K+M-1-W)\j(s)\\
&& + \sum_{s=1}^{N_5-1} \left(\bj(s)P_R\j(s+1)+\bj(s+1)P_L\j(s)\right)
\nonumber\\
&& - \ m_0\left(\bj(N_5)P_R\j(1)+\bj(1)P_L\j(N_5)\right)\ ,\nonumber
\end{eqnarray}
where $\j$ is the five-dimensional fermion field $\j(x,s)$, $s=1,\dots,N_5$.
In Eq.~(\ref{Sdwf}), only
the dependence on the fifth coordinate is made explicit.
The mass term couples the fields on opposite boundaries.

Domain-wall fermions are not exactly massless for finite $N_5$ when $m_0=0$.
The mass induced by a finite fifth direction, usually referred to as the residual mass,
is reminiscent of the additive mass renormalization of Wilson fermions.
However, the residual mass vanishes in the limit $N_5\to\infty$,
which we will take {\em before} the continuum limit.
Following this order of limits,
the mass term introduced in Eq.~(\ref{Sdwf}) renormalizes multiplicatively.
Thus, the complications of the additive mass renormalization
of the Wilson case, that we encountered in Sec.~\ref{Wilson}, are avoided.

Our aim in this subsection is to recast the argument given in Sec.~\ref{Wilson} in terms of the
domain-wall formulation of the lattice regularized theory.   The first step is to prove an analogous
result to Eq.~(\ref{Zlim}), thus rederiving the theorem of Ref.~\cite{SSt} in terms of domain-wall fermions.
For this, we need to define an axial transformation.   We take $N_5=2K$ even, and define the axial transformation as \cite{YSdwf2}
\begin{eqnarray}
\label{axial}
\d\j(s)&=&e^{i\h}\j(s)\ ,\qquad\ \ \!\d\bj(s)=\bj(s)e^{-i\h} \ ,\qquad 1\le s\le K\ ,\\
\
\d\j(s)&=&e^{-i\h}\j(s)\ ,\qquad\d\bj(s)=\bj(s)e^{i\h}\ ,\quad K+1\le s\le 2K\ .\nonumber
\end{eqnarray}
Following Ref.~\cite{YSdwf2}, we define the five-dimensional currents
\begin{eqnarray}
\label{currents}
j_\m(x,s)&=&\half\left(\bj(x,s)(1+\g_\m)U_\m(x)\j(x+\m,s)-\bj(x+\m,s)(1-\g_\m)U^\dagger_\m(x)\j(x,s)\right)\ ,\nonumber\\
j_5(x,s)&=&\bj(x,s)P_R\j(x,s+1)-\bj(x,s+1)P_L\j(x,s)\ .
\end{eqnarray}
The four-dimensional axial current corresponding to the axial transformation~(\ref{axial})
is \begin{equation}
\label{axialcurr}
j^A_\m(x)=-\sum_{s=1}^K j_\m(x,s)+\sum_{s=K+1}^{2K} j_\m(x,s)\ .
\end{equation}
It satisfies the Ward--Takahashi identity
\begin{equation}
\label{WI}
\partial^-_\m j^A_\m=2j_5(K)+2m(\bj(2K)P_R\j(1)-\bj(1)P_L\j(2K))\ .
\end{equation}

Analogous to Eq.~(\ref{Wth}), we can now introduce two angles,
through the combinations $S_W(\th_W)$ and $S_m(\th_m)$, where
\begin{eqnarray}
\label{DWFthth}
S_W(\th_W) &=& e^{i\theta_W}\bj(K)P_R\j(K+1)+e^{-i\theta_W}\bj(K+1)P_L\j(K)\ , \\
S_m(\th_m) &=& -m_0\left(e^{-i\theta_m}\bj(2K)P_R\j(1)+e^{i\theta_m}\bj(1)P_L\j(2K)\right)\ .\nonumber
\end{eqnarray}
$S_W(\th_W)$ replaces the $s=K$ term on the second line of Eq.~(\ref{Sdwf}),
and $S_m(\th_m)$ replaces the mass term (third line) in Eq.~(\ref{Sdwf}).
Once again, under an axial transformation (Eq.~(\ref{axial})), $\th_{m,W}\to\th_{m,W}+2\h$, and hence
only the difference $\th_F=\th_W-\th_m$ is physical.

Slightly generalizing the discussion of the previous subsection,
here we will keep both $\th_W$ and $\th_m$ arbitrary.
If we now differentiate the fermion partition function with respect
to $\th_W$, the result is $\svev{\tj_5(\th_W)}$,
where $\svev{\cdot}$ indicates integration over the fermion fields,
and we have defined
\begin{equation}
\label{dSdwf}
\tj_5(\th_W) = e^{i\th_W}\bj(K)P_R\j(K+1) - e^{-i\th_W}\bj(K+1)P_L\j(K)\ ,
\end{equation}
We will prove that in the theory with a non-zero $\th_W$,
the continuum limit of $\svev{\tj_5(\th_W)}$ yields the axial anomaly.
By integrating with respect to $\th_W$, it then follows that
\begin{equation}
\label{Zththdwf}
Z(\th,\th_W,\th_m)=Z(\th+NT\th_W,0,\th_m)\ ,
\end{equation}
where now the path integral is defined with the domain-wall fermion action instead
of the Wilson fermion action, and we have again allowed
for $N$ Dirac fermions in an \irrep\ with index $T$.
Equation~(\ref{Zththdwf}) generalizes Eq.~(\ref{Zlim})
of the preceding subsection.

The proof turns out to be quite straightforward.
Let $G(\th_W,\th_m)$ be the inverse of the domain-wall Dirac operator
$D(\th_W,\th_m)$, with angles $\th_W$ and $\th_m$ introduced as in
Eq.~(\ref{DWFthth}).
Using Eq.~(\ref{dSdwf}), and writing $\tj_5(\th_W)=\bj J_5(\th_W)\j$, we have
\begin{equation}
\label{proof}
\svev{\tj_5(\th_W)}=-\Tr\Big(J_5(\th_W)G(\th_W,\th_m)\Big)=-\Tr\Big(J_5(0)G(0,\th_m-\th_W)\Big)\ ,
\end{equation}
where in the second step we used the axial transformation~(\ref{axial})
with $\h=\th_W/2$ to move the angle $\th_W$ to the mass term.
We now take the limit $K\to\infty$, in which the propagator in Eq.~(\ref{proof}) becomes translationally invariant
in the fifth dimension. In particular, the propagator becomes independent
of the boundaries, and thus of $m$ and $\theta_m$
(or $\th_m-\th_W$ after the axial rotation).
It follows that $\svev{\tj_5(\th_W)}=\svev{\tj_5(0)}$
for any $\th_W$ and $\th_m$, and the anomaly is recovered as in
Ref.~\cite{YSdwf1}.

With the domain-wall equivalent of Eq.~(\ref{Zlim}) in hand, we now return to the equivalence between
one Dirac fermion in a real \irrep\ of the gauge group and two Majorana fermions, in the domain-wall regularization.
As we will see, the argument follows similar steps as that for
the Wilson-fermion case given in Sec.~\ref{Wilson}.

We begin by mapping the action~(\ref{Sdwf}) into an
action for two Majorana fermions, denoted as $\J_i$, $i=1,2$. We again make use of Eq.~(\ref{Diracreal}), but now with
a Majorana condition adapted for domain-wall fermions.   Analogous to
Eq.~(\ref{majcond}),  we will require that \cite{DBKMS}
\begin{equation}
\label{bJdwf}
\bJ=(R_5\J)^TCS\ ,
\end{equation}
with $S$ and $C$ as in Sec.~\ref{maj},
and $R_5$ a reflection in the fifth direction:
\begin{equation}
\label{R}
R_5\J(x,s)=\J(x,N_5-s+1)\ .
\end{equation}
The reason for adding the reflection is that charge conjugation (in four dimensions)
interchanges left- and right-handed fermions.   Here the right- and left-handed modes
emerge near  the boundaries $s=1$ and $s=N_5$, respectively, and they need to be
explicitly interchanged to match the four-dimensional picture.    The domain-wall
fermion action~(\ref{Sdwf}) in terms of two massless Majorana fermions $\J_{1,2}$ defined
by
\begin{eqnarray}
  \J_{L,1}(s) &=& \j_L(s) \ ,
\label{Psidwf}\\
  \J_{R,1}(s) &=& R_5SC\,\bj^T_L(s)=SC\,\bj^T_L(N_5-s+1) \ ,
\NON
  \J_{L,2}(s) &=& R_5SC\,\bj^T_R(s)= SC\,\bj^T_R(N_5-s+1)\ ,
\NON
  \J_{R,2}(s) &=& \j_R(s) \ ,
\nonumber
\end{eqnarray}
is then given, for $m_0=0$, by
\begin{eqnarray}
\label{Sdwfmaj}
S_{\rm Maj}&=&\half\sum_{s=1}^{N_5}\J^T(N_5-s+1)CSD_K\J(s)\\
&& +\ \half\sum_{s=1}^{N_5} \J^T(N_5-s+1)CS \s_1 (M-W-1)\J(s)\nonumber\\
&& +\ \half\sum_{s=1}^{N_5-1} \left(\J^T(N_5-s+1)CS \s_1 P_R\J(s+1)
                            +\J^T(N_5-s)CS \s_1 P_L\J(s)\right)\ ,
\nonumber
\end{eqnarray}
where $\s_1$ is again the first Pauli matrix acting on the flavor index $i=1,2$ of $\J_i$.
Using Eq.~(\ref{R}) and Eq.~(\ref{Psidwf}),
the Majorana form of Eq.~(\ref{DWFthth}) is
\begin{eqnarray}
\label{DWFththmaj}
S_W(\th_W) &=& \half\left(e^{i\th_W}\J^T_R(K+1)SC\s_1\J_R(K+1)+e^{-i\th_W}\J^T_L(K)SC\s_1\J_L(K)\right)\,, \hspace{6ex} \\
S_m(\th_m) &=& -\frac{m_0}{2}\left(e^{-i\th_m}\J^T_R(1)SC\s_1\J_R(1)+e^{i\th_m}\J^T_L(2K)SC\s_1\J_L(2K)\right)\,.\nonumber
\end{eqnarray}
$S_m(\th_m)$ gets added to the massless Majorana domain-wall action~(\ref{Sdwfmaj}),
while $S_W(\th_W)$ replaces the $s=K$ term
on the third line of Eq.~(\ref{Sdwfmaj}).

As in Sec.~\ref{Wilson}, the flavor matrix $\s_1$ in Eq.~(\ref{DWFththmaj}) can be rotated into $\s_3$.   If we then
perform a phase transformation\footnote{Again, this phase transformation is not anomalous
on the lattice.}
\begin{eqnarray}
\label{Psi2phase}
\J_2(x,s)&\to& i\J_2(x,s)\ ,\qquad 1\le s\le K\ ,\\
\J_2(x,s)&\to& -i\J_2(x,s)\ ,\qquad K+1\le s\le 2K\ ,\nonumber
\end{eqnarray}
on the Majorana field $\J_2$, while leaving $\J_1$ alone, this rotates
$\s_3$ into the identity matrix in flavor space.
The end result is that $\s_1$ is removed from Eqs.~(\ref{Sdwfmaj}) and~(\ref{DWFththmaj}) (while leaving the kinetic term invariant), thus
proving that the theory has two Majorana
fermions with equal positive mass $m$ and the same $\th$ angle as the Dirac theory.
Again, using that $\pf^2(\ca)=\det(\ca)$ for any antisymmetric $\ca$, we conclude that the Majorana
measure is identical to the Dirac measure.

\subsection{\label{smr} Summary}
We summarize the main results of this section.  The starting point
is a lattice-regularized theory with Wilson or domain-wall fermions,
and with chiral angles $\th_m$ and $\th_W$ introduced in
Eqs.~(\ref{Wth}) or~(\ref{DWFthth}), respectively.\footnote{
  Or, to be more precise,
  in Eq.~(\ref{Diracren}) in the case of Wilson fermions.}
We also allow for
an explicit topological term in the gauge action, with angle $\th$
(see Eq.~(\ref{lag})).

Consider first the case of $N$ identical Dirac fermions.
As first observed in Ref.~\cite{SSt},
in the continuum limit an additional vacuum angle
\begin{equation}
\label{thind}
  \thind = NT\th_W \ ,
\end{equation}
is induced by the fermions.
Introducing the ``shifted'' angle
\begin{equation}
\label{thetatot}
  \thshft = \th + \thind\ ,
\end{equation}
the operational meaning of this statement is that
all observables will be reproduced in the continuum limit
if we set $\th_W=0$, and, at the same time, replace $\th$ by $\thshft$
as the angle multiplying the explicit topological term in the
(lattice) lagrangian.
As for the phase of the fermion mass matrix,
we trivially have $\a=NT\th_m$ (recall Eq.~(\ref{cm})).
Substituting this into Eq.~(\ref{theff}) we conclude that,
after integrating out the fermions, the effective
vacuum angle in the gauge field measure is
\begin{equation}
\label{thefflatt}
  \theff = \thshft-\a = \th + NT(\th_W-\th_m) \ .
\end{equation}
In the case of $N_{maj}$ identical Majorana fermions,
the same result holds, with $N=N_{maj}/2$.

The interesting case is an even number $2N$ of Majorana fermions,
which we have shown to be equivalent to $N$ Dirac fermions, as they should be.
This has resolved the apparent paradox described in Sec.~\ref{intro}.
We conclude this section by summarizing the result in the case of
a single Dirac fermion, $N=1$.

The key observation is that, after the transition from a Dirac fermion
to two Majorana fermions, the mass term and the Wilson term
(or its domain-wall fermion counterpart) are proportional to the same matrix in flavor space.
As we have shown, by a sequence of non-anomalous lattice transformations
(meaning that the jacobian of each lattice transformation is equal to one),
we may bring the two Majorana fermions to a diagonal form,
with the same phases as for the original Dirac fermion
(see, \eg, Eq.~(\ref{finalWmaj}) for the Wilson case).

Alternatively, we may elect to apply only SU(2) transformations
to the Majorana fermions.
These can bring the Wilson and mass terms, that originally point
in the $\s_1$ direction in flavor space, first into the $\s_3$ direction,
and then into the $i\g_5$ direction (see Eq.~(\ref{Js2})).
In this situation we again obtain two identical Majorana fermions,
except with new phases that are shifted by the same amount,
$\th'_W=\th_W+\p/2$ and $\th'_m=\th_m+\p/2$.
In the continuum limit the explicit topological phase becomes
$\thshft' = \th + T\th'_W=\th+T(\th_W+\p/2)$.
Because the difference $\th'_W-\th'_m=\th_W-\th_m$ is unchanged,
when we substitute the new phases into Eq.~(\ref{thefflatt}) we see that
the effective vacuum angle $\theff$ is unchanged as well.

\section{\label{dirac} Vacuum angle and the chiral condensate: complex \bf{\textit{irrep}}}
Our non-perturbative study in the previous section has implications
for the chiral expansion of fermions in a real \irrep,
and, in particular, for the interplay between the vacuum angle
and the U(1) phase of the fermion mass matrix within the chiral expansion.
These will be discussed in Sec.~\ref{vacreal} below.
As a preparatory step, in this section we review the role
of the vacuum angle in the more familiar case of fermions in a complex \irrep.
We first consider the chiral condensate in the underlying theory
in Sec.~\ref{diracmic}, paying special attention to its U(1) phase
in the light of the results of the previous section.
In Sec.~\ref{dirachpt} we then discuss how the same features are reproduced
in the effective theory, \ie, in chiral perturbation theory.

\subsection{\label{diracmic} Microscopic theory}
We begin with a continuum derivation.
Starting from Eqs.~(\ref{Z}),~(\ref{lag}) and~(\ref{cm}),
the left-handed and right-handed fermion condensates are defined by
\begin{subequations}
\label{cond}
\begin{eqnarray}
  \S_{L,ij} &=& \svev{\bj_j P_L \j_i} \ = \
  -\frac{1}{V}\frac{\partial \log Z}{\partial \cm^*_{ij}}\ ,
\label{condL}\\
  \S_{R,ij} &=& \svev{\bj_j P_R \j_i} \ = \
  -\frac{1}{V}\frac{\partial \log Z}{\partial \cm_{ji}} \ ,
\label{condR}
\end{eqnarray}
\end{subequations}
where $V$ is the volume, and $i,j=1,\ldots,N$ are flavor indices.
Standard steps using the identity
\begin{equation}
  \Big(\Sl{D} + m(\O^\dagger P_L + \O P_R) \Big)
  \Big(-\Sl{D} + m(\O P_L + \O^\dagger P_R) \Big)
  = -\Sl{D}^2 + m^2 \ .
\label{Dsq}
\end{equation}
give rise to the expressions
\begin{eqnarray}
  \S_L &=& -(a_1-a_5) \O \ ,
\label{SigL}\\
  \S_R &=& -(a_1+a_5) \O^\dagger \ ,
\label{SigR}
\end{eqnarray}
where
\begin{eqnarray}
  a_1 &=& \frac{m}{2V}
  \svev{\Tr \left[\Big(-\Sl{D}^2 + m^2\Big)^{-1} \right] } \ ,
\label{g0cond}\\
  a_5 &=&  \rule{0ex}{4ex} \frac{m}{2V}
  \svev{\Tr \left[\g_5 \Big(-\Sl{D}^2 + m^2\Big)^{-1} \right] } \ .
\label{g5cond}
\end{eqnarray}
The $\Tr$ symbol indicates a trace over spacetime, color
and Dirac indices.\footnote{When the Dirac operator occurs inside the $\Tr$ symbol,
by convention it carries no flavor indices.}
By applying a parity transformation we may express these
quantities more explicitly as
\begin{eqnarray}
  a_1 &=& \frac{m}{2V} \int \cd A\, \tm(A) \cos(\theff Q)
  \,\Tr\! \left[\Big(-\Sl{D}^2 + m^2\Big)^{-1} \right] \ ,
\label{vala1}\\
  a_5 &=& -\frac{im}{2V} \int \cd A\, \tm(A) \sin(\theff Q)
  \,\Tr\! \left[\g_5\Big(-\Sl{D}^2 + m^2\Big)^{-1} \right] \ .
\label{vala5}
\end{eqnarray}
It follows that $a_1$ is real, while $a_5$ is imaginary.
Both $a_1$ and $a_5$ are functions of $\theff$, defined in Eq.~(\ref{theff}).
Introducing
\begin{equation}
  z = a_1 - a_5\ ,
\label{z}
\end{equation}
we arrive at
\begin{equation}
  \S_L = \S_R^\dagger =  -z(\theff)e^{i\a/(NT)}\tO=-\left[z(\theff)e^{-i\theff/(NT)}\right]e^{i\th/(NT)}\tO \ .
\label{Sigz}
\end{equation}
In the special case $\theff=\th-\a=0$,
$a_5$ vanishes while $a_1=r$ is real positive.
Hence, in that case, $z=r>0$, and
\begin{equation}
\label{Sigr}
  \S_L = -r\O = -r\, e^{i\a/(NT)} \tO=-r\, e^{i\th/(NT)} \tO \ .
\end{equation}
Finally, in the limit $m\to 0$ we recover the
Banks--Casher relation,
\begin{equation}
\label{BC}
  r = \frac{\p}{2}\,\r(0) \ ,
\end{equation}
where $\r(\l)$ is the spectral density of the massless Dirac operator.

Returning to the general case of Eq.~(\ref{Sigz}) we see that
the orientation of the condensate is determined by the
``normalized'' mass matrix $\cm/m$ and by $\theff$.
In retrospect, this pattern is a consequence of Eq.~(\ref{cond}),
which defines the condensates via derivatives of the partition function with respect to
the mass matrix, together with the fact that the partition function itself
is invariant under non-abelian chiral rotations of the mass matrix,
and depends on $\th$ (or $\thshft$) and $\a$ through their difference $\theff$ only,
as we proved rigorously in Sec.~\ref{latt}
(see, in particular, Eq.~(\ref{thefflatt})).  These are the only features
of the condensate that we will need in the following.

\subsection{\label{dirachpt} Effective low-energy theory}
We now turn to the effective theory for the Nambu--Goldstone pions associated with the
spontaneous breaking of chiral symmetry.
As noted above, at this stage the discussion is restricted to QCD-like theories
in which the fermions belong to a complex \irrep.
The dynamical effective field  is
\begin{equation}
\S(x) = \S_0 U(x)\ ,\qquad  U(x) = \exp(i\sqrt{2}\P(x)/f) \ , \qquad
  \P(x)=\prod_{a=1}^{\Nf^2-1}\P_a(x)T_a \ ,
\label{Upion}
\end{equation}
where $U(x)$ is the $\SU(N)$ valued pion field, and $\S_0\in \Uone$ is a constant phase factor.\footnote{
  Any constant $\SU(N)$-valued part of $\S$ can be absorbed
  into the pion field. $\S_0$ may be regarded as a remnant of the $\h'$ field
  (see, for instance, Refs.~\cite{EW,VV}).
}
The leading-order potential is
\begin{equation}
\label{Vcl}
  V = - \frac{f^2 B}{2} \tr(\cm^\dagger \S + \S^\dagger \cm) \ ,
\end{equation}
where we recall that $\cm=me^{i\a/(NT)}\tO$, with $\tO\in \SU(N)$.
We remind the reader that the product $Bm$
is renormalization-group invariant, and depends only
on the chiral-limit value of the condensate.\footnote{
  In particular, the leading-order chiral lagrangian is insensitive
  to the quadratic divergence proportional to $m/a^2$ that
  is present in the bare lattice condensate away from the chiral limit
  in any fermion formulation.
}

As we have seen in Sec.~\ref{diracrev}, the partition function of the microscopic theory
depends on $\a$ and $\th$ only through their difference
$\theff = \th - \a$, and the same must thus be true in the effective theory:
the lagrangian of the effective theory must be a function
of $\theff$ only, order by order in the chiral expansion, starting with the
tree-level potential $V$.  Evidently, $V$ will be a function of only $\theff$ if we set
\begin{equation}
\label{Sig0}
  \S_0 = e^{i\th/(NT)} \ .
\end{equation}
In App.~\ref{theffchpt} we use the power counting and the symmetries
of the effective theory to prove that Eq.~(\ref{Sig0})
provides the unique solution to the requirement that
the tree-level potential~(\ref{Vcl}) depends on $\a$ and $\th$ only
through their difference $\theff$.  We also prove that a similar statement
applies to the next-to-leading order lagrangian.

In the effective theory, the tree-level condensate now takes the form
\begin{equation}
\label{Sigth}
  \S_L =\frac{\partial V}{\partial \cm^*}  \bigg|_{U=U_0}= -\frac{f^2 B}{2}\, e^{i\th/(NT)} U_0 \ ,
\end{equation}
where $U_0 \in \SU(N)$ is the global minimum of the potential.
For this to be consistent with Eq.~(\ref{Sigz}), the global minimum
$U_0$ must be equal to $U_n = e^{2\p in/N} \tO$, for some $0\le n < N$, as we will
see next.
Substituting $U_n$ into Eq.~(\ref{Vcl}) gives
\begin{equation}
\label{Vmin}
  V(U_n)=-f^2 BNm \cos(\theff/(NT)+2\p n/N) \ .
\end{equation}
In App.~\ref{proofSigLO} we prove that the
global minimum is reached when $\theff+2\p nT$ is closest to zero.
Denoting the corresponding value of $n$ by $n(\theff)$,
the tree-level condensate is thus
\begin{equation}
\label{SigLO}
  \S_L = -\frac{f^2 B}{2} e^{i(\th/(NT)+2\p n(\theff)/N)} \tO \ .
\end{equation}
This result for $\S_L$ is consistent with Eq.~(\ref{Sigz}), and thus demonstrates the
need to introduce the constant $\Uone$-valued phase $\S_0$ into the effective theory.
Without $\S_0$, the effective theory would have led to a value for $\S_L$ in
$\SU(N)$.   This would have been inconsistent, as, for example, can be seen in
the case $\th=\a\ne 0$, by comparison with Eq.~(\ref{Sigr}).

We comment that
in exceptional cases there is a competition between the leading-
and next-to-leading order potentials \cite{Smilga,HS}.
In that case the discussion leading to Eq.~(\ref{SigLO}) does not apply.
But the functional form of Eq.~(\ref{SigLO}) remains valid:
it must remain true that $\S_L$ is oriented in
the direction of $e^{i(\th/(NT)+2\p n/N)} \tO$ for some $n$,
where again $n$ depends on $\theff$ only, as can again be seen
by comparison with Eq.~(\ref{Sigz}).

\section{\label{vacreal} Vacuum angle and the chiral condensate: real \bf{\textit{irrep}}}
In this section we turn to real \irreps.
In Sec.~\ref{real} we discuss the condensate,
and elaborate on the differences between the complex case
(discussed in Sec.~\ref{dirac}) and the real case.
We deal separately with a stand-alone theory of Majorana fermions,
and with a theory of $2N$ Majorana fermions that was obtained
by reformulating a theory of $N$ Dirac fermions, where the apparent paradox
described in the introduction arises.
We then discuss the implications for the chiral effective theory.
In Sec.~\ref{chpt} we give a diagrammatic proof that,
when $\theff$ is held fixed, different orientations
of the mass matrix give rise to same physical observables.

\subsection{\label{real} The condensate for a real \textbf{\emph{irrep}}}
We begin with a general theory of $\Nmaj$ Majorana fermions,
where $\Nmaj$ can be both even or odd.  Allowing $N=\Nmaj/2$
to be half-integer in Eq.~(\ref{transfreal}),
the flavor symmetry of the massless theory is $\SU(\Nmaj)$,
which we will assume to be spontaneously broken to $\SO(\Nmaj)$.
We will consider a mass term of the general form
\begin{equation}
\label{genmassmaj}
\half\bJ(\cm^\dagger P_L+\cm P_R)\J\ ,
\end{equation}
where now
\begin{equation}
\label{majmass}
\cm = \cm^T = m\O = m e^{2i\a/(\Nmaj T)} \tO \ , \qquad \tO \in \SU(\Nmaj) \ ,
\end{equation}
and we assume $m>0$.  Formally, the fermion path integral is a pfaffian.
However, as we have seen in Sec.~\ref{ambg}, the phase of this pfaffian
is ambiguous in the continuum.  The rigorous solution to this problem
is to define the pfaffian via a lattice regularization.
For the mass matrix in Eq.~(\ref{majmass}), this gives rise
to the following relations in the continuum limit
\begin{equation}
\label{PfF}
  \pf(\Sl{D} + \cm^\dagger P_L + \cm P_R)
  = e^{i\a Q}\, \pf^{\Nmaj}(\Sl{D}+m)
  = e^{i\a Q}\, \det^{\Nmaj/2}(\Sl{D}+m)\ .
\end{equation}
The second equality implies that $\pf(\Sl{D}+m)$ is strictly positive,
as follows from App.~\ref{pospf}.
One way to derive Eq.~(\ref{PfF}) is to start from a lattice theory of
domain-wall Majorana fermions with $\th_W=0$ and $\th_m=2\a/(\Nmaj T)$,
and take the continuum limit.
Defining $\S_L$ and $\S_R$ as in Eqs.~(\ref{condL}) and~(\ref{condR}),
but replacing $\j\to\J$ and $\bj\to\bJ$, the rest of the discussion
of Sec.~\ref{diracmic} carries over.\footnote{
  The definition of parity is somewhat more subtle with Majorana fermions, see
Ref.~\cite{GSHiggs2}.}

We next consider the case where $N$ Dirac fermions are traded
with $2N$ Majorana fermions. In the initial Dirac-fermion lattice formulation
we again set $\th_W=0$.  As follows from Sec.~\ref{latt},
this choice implies that $\thshft=\th$,
and thus the angle $\th$ that multiplies the lattice-discretized
topological term in the gauge action is set to the same value
as in the target continuum theory.  As usual, the $\Uone$ phase of the
lattice mass matrix is the same as in the continuum, $\th_m=\a/(NT)$.

The key point is that the values of $\th_m$ and $\th_W$
in any equivalent Majorana formulation are constrained by their values
in the initial Dirac formulation, and, in particular, by the
choice $\th_W=0$ we have initially made.  The basic transition
to Majorana fermions (using Eq.~(\ref{Diracreal}) in the Wilson case,
or Eq.~(\ref{Psidwf}) in the domain-wall case) gives rise to a mass term
and a (generalized) Wilson term that are both oriented in the direction
of the matrix $J_S$ of Eq.~(\ref{Js}).  In itself, $J_S$ has an axial $\Uone$ phase
of $\p/2$. 
As a result, the phases of the mass term
and the (generalized) Wilson term both get shifted by $\p/2$,
becoming $\th'_m = \a/(NT)+\p/2$, and $\th'_W = \p/2$.
In the continuum limit, the new phase of the mass matrix
is $\a' = NT\th'_m = \a+NT\p/2$.  The phase $\th'_W$ gets traded
with an additional vacuum angle, so that the new vacuum angle
is $\th'=\thshft=\th+NT\p/2$.  As expected, both phases were shifted
by the same amount, so that the effective vacuum angle,
which is their difference, is unchanged, $\theff = \th-\a = \th'-\a'$.

Alternatively, we may perform an additional (non-anomalous)
lattice transformation that brings back the phases to their original values,
$\th_m = \a/(NT)$ and $\th_W=0$, so that $\thshft=\th$
(for the Wilson case, see Eq.~(\ref{finalWmaj})).
Once again, $\theff$ is unchanged.

We next turn to the chiral effective theory, focusing on the case $\Nmaj=2N$,
with the mass matrix $\cm$ of Eq.~(\ref{majmass}).  The non-linear field $\S$
is now an element of the coset $\SU(2N)/\SO(2N)$.
It is symmetric, $\S^T=\S$, and transforms
as $\S\to h\S h^T$ under the chiral transformation~(\ref{transfreal}),
just like $\cm$ (when elevated to a spurion).
Instead of Eqs.~(\ref{Upion}) and~(\ref{Sig0}), which we had in the case
of a complex \irrep, the coset field for a real \irrep\ is parametrized as
\begin{equation}
\label{Sigreal}
  \S(x) = U(x)^T \S_0 = \S_0 U(x)\ ,
\end{equation}
where now
\begin{equation}
\label{Sig0J}
  \S_0 = e^{i\tilde\th/(NT)} J \ ,
\end{equation}
and where $J$ is a real symmetric $\SO(2N)$ matrix.
Once again, the phase $\tilde\th$ is to be chosen so that the chiral theory
is a function of $\theff$ only.  We will discuss examples of this shortly.
Equations~(\ref{Sigreal}) and~(\ref{Sig0J}) provide a generalization
of the results of Ref.~\cite{BL}, where the role of the U(1) phase was not
discussed, and of Ref.~\cite{tworeps}, where the discussion was limited
to $\theta=\alpha=0$, and $J=\id_{2N}$.

For simplicity, in the rest of this section we again set $\tO=1$
in Eq.~(\ref{majmass}).\footnote{%
  The generalization to arbitrary $\tO$ is similar to Sec.~\ref{dirac}.
}
Let us consider the construction of the chiral theory in the case
we have just discussed, where $N$ Dirac fermions get traded
with $2N$ Majorana fermions.  In the initial Dirac formulation we take
the mass matrix to be $me^{i\a/(NT)}\id_N$,
and we allow for an arbitrary vacuum angle $\th$.
After the transition to the Majorana formulation,
the mass matrix is $\cm=me^{i\a/(NT)} J_S$, which
is equivalent to a $\Uone$ phase
$\a'/(NT)=\a/(NT)+\p/2$.  Correspondingly, the vacuum angle of
the continuum-limit theory becomes $\th'=\th+NT\p/2$.
A possible choice for $\S_0$ is $e^{i\th'/(NT)} \id_{2N}$.
An alternative, equivalent choice, which involves the same U(1) phase,
is $\S_0=e^{i\th/(NT)} J_S$.  For the latter choice, the factors of $J_S$
cancel out between the mass matrix and the non-linear field
when the latter is expanded in terms of the pion field.
Studying the classical solution as we did in Sec.~\ref{dirachpt},
we similarly find that the expectation value of the pion field $U(x)$
is a $Z_{2N}$ element which again depends only on $\theff$.

The situation is similar if we apply an $\SU(2N)$ transformation
that rotates the Majorana mass matrix to
$\cm=me^{i(\a/(NT)+\p/2)} \id_{2N} \ = me^{i\a'/(NT)} \id_{2N}$
(this corresponds to $J_S^{\rm rot}$ of Eq.~(\ref{Js2})).
If we choose to apply the same $\SU(2N)$ rotation to $\S_0$, it becomes
$\S_0=e^{i(\th/(NT)+\p/2)} \id_{2N}=e^{i\th'/(NT)} \id_{2N}$.
Finally, if in the lattice-regularized theory we have applied
a further U(1) axial transformation that simultaneously brings the mass matrix
to $\cm=me^{i\a/(NT)} \id_{2N}$, and the (shifted) vacuum angle
of the continuum-limit theory back to $\thshft=\th$, then in the chiral theory
we can correspondingly set $\S_0=e^{i\th/(NT)} \id_{2N}$.
In all of these examples, the constant mode of the pion field $U(x)$
will be a $Z_{2N}$ element determined by $\theff$ only.

\subsection{\label{chpt} Chiral expansion for a real \textbf{\emph{irrep}}}
In the case of a complex \irrep, studied in Sec.~\ref{dirac}, we have
demonstrated that the condensate can be expressed as a function
of $\th$ and $\theff$ via Eq.~(\ref{Sigz}).
We then determined the $\th$ dependence of the chiral lagrangian
by requiring that the effective theory reproduce this result.
When we expand the chiral lagrangian around the classical solution
in terms of the pion field, the expansion is then manifestly a function
of $\theff$ only, and not of $\th$ and $\a$ separately.
It follows that physical observables, such as the decay constant
and the pion mass, depend only on $\theff$ as well.

In the case of a real \irrep, we again expect that the chiral expansion
for any physical observable will depend on $\a$ and $\th$ only through
their difference $\theff$.  However, establishing this result is now
more subtle.  Let us consider two simple examples, both of
which can be parametrized as $\cm=mJ$, $\S_0=J$, for the same $J$.
The two cases are then defined by taking $J=J_S$,
for which $\a/(NT)=\th/(NT)=\p/2$, or $J=\id_{2N}$,
for which $\a=\th=0$.  Notice that $\theff=0$ in both cases.
Now, using Eq.~(\ref{Sigreal}), and noting that in both cases $J^2=\id_{2N}$,
it is easy to see that $J$ drops out of the product $\S^\dagger(x)\cm$.
However, unlike in the case of a complex \irrep, this does not
immediately imply that the perturbative expansion is independent of
the choice of $J$.  The reason is the constraints imposed on
the pion field: this field is hermitian, traceless, and satisfies
\begin{equation}
\label{constraint}
  \p=J\p^T J\ .
\end{equation}
Thus, even though $J$ drops out of the tree-level lagrangian,
the pion field still depends on it, through the above constraint,
and the pion propagator \cite{BL,tworeps}
\begin{equation}
\label{pionprop}
  \svev{\p_{ij}(x)\p_{k\ell}(y)}
  = \int\frac{d^4p}{(2\pi)^4}\, \frac{e^{ip(x-y)}}{p^2+M^2}
  \left(\frac{1}{2} \left(\delta_{i\ell}\delta_{jk} + J_{ik}J_{j\ell}\right)
  - \frac{1}{2N}\,\delta_{ij}\delta_{k\ell}\right)\ ,
\end{equation}
depends on the choice of $J$ explicitly.

Let us consider the case $N=1$.
For $J=J_S$, and choosing a basis where $J_S=\s_3$,
the constraints translate into
$\p_{11}=\p_{11}^*=-\p_{22}$, and $\p_{12}=-\p_{12}^*=-\p_{21}$.
For $J=\id_2$, the diagonal elements remain the same as before,
whereas for the off-diagonal elements we have $\p_{12}=\p_{12}^*=\p_{21}$.
Stated differently, for $J=\s_3$ the expansion of the pion field
is $\p=\p_3\s_3+\p_2\s_2$, whereas for $J=\id_2$
it is $\p=\p_3\s_3+\p_1\s_1$.  The tensor structure of the non-vanishing
propagators is
\begin{eqnarray}
\label{propexample}
\svev{\p_{11}(x)\p_{11}(y)}:\quad
\frac{1}{2} \left(\delta_{11}\delta_{11} + J_{11}J_{11}\right)
  - \frac{1}{2}\,\delta_{11}\delta_{11}&=&\frac{1}{2}\ ,
  \qquad J=\s_3,\id_2\ ,\\
\svev{\p_{12}(x)\p_{12}(y)}:\quad
\frac{1}{2} \left(\delta_{12}\delta_{12} + J_{11}J_{22}\right)
  - \frac{1}{2}\,\delta_{12}\delta_{12}&=& \rule{0ex}{5ex}
  \left\{ \begin{array}{c} -\half \ , \qquad J=\s_3 \ , \\
                           \half \ , \qquad J=\id_2 \ .
          \end{array} \right.\nonumber
\end{eqnarray}
Using a hat to distinguish the pion field for the case $J=\s_3$,
we see that it will transform into the pion field of the $J=\id_2$ case
if we substitute
\begin{equation}
\label{fieldredef}
\hp_{11}=\p_{11}\ ,\qquad \hp_{12}=i\p_{12}\ ,
\end{equation}
which corresponds to the replacement of $\s_1$ by $\s_2$
in the expansion of the pion field.  Of course, non-perturbatively,
the redefinition~(\ref{fieldredef}) is not allowed,
but in (chiral) perturbation theory the only question
is whether it leads to the same order-by-order diagrammatic expansion
for any correlation function with a prescribed set of external pion legs.
We will now prove that
\begin{eqnarray}
\label{lemma}
&&\svev{\hp_{11}^{(1)}(x_1)\dots \hp_{11}^{(m)}(x_m)\hp_{12}^{(1)}(y_1)\dots\hp_{12}^{(n)}(y_n)}\\
&&\hspace{2cm}=i^n\svev{\p_{11}^{(1)}(x_1)\dots \p_{11}^{(m)}(x_m)\p_{12}^{(1)}(y_1)\dots\p_{12}^{(n)}(y_n)}\ ,
\nonumber
\end{eqnarray}
to all orders in chiral perturbation theory, for any $m$ and $n$.

A vertex with $k$ $\p_{12}$ lines attached to it also changes by a factor $i^k$ after
the field redefinition (note that $k$ is always even, so that taking $i$ or $-i$ does not matter).
Also, for any diagram,
the number of $\p_{12}$ external lines $n$, the number of $\p_{12}$ propagators $p$
and the number $v_k$ of vertices with $k$ $\p_{12}$ lines attached to it are related by
\begin{equation}
\label{12comb}
2p=n+\sum_k k v_k\ .
\end{equation}
It follows from this relation that, for all diagrams, the field redefinition~(\ref{fieldredef}) indeed
leads to the factor $i^n$ in Eq.~(\ref{lemma}), thus proving this result.
Each $\p_{12}$ propagator flips its sign, and $p$ such propagators thus
lead to a factor $(-1)^p=i^{2p}$.
In addition, the diagram changes by a factor $i^{\sum_k k v_k}$ because of
the $v_k$ vertices with $k$ $\p_{12}$ lines, and thus the diagram changes by a total
factor $i^{2p+\sum_k k v_k}=i^n$, using Eq.~(\ref{12comb}).   Here we also used that all terms in the
exponent are even (and, thus, $n$ is even as well).

Next, we discuss the general case of $N$ Dirac fermions in a real \irrep,
comparing the cases $J=J_S$, with $J_S$ in Eq.~(\ref{Js}), and $J=\id_{2N}$.
The matrix $J_S$ can now be brought onto a form in which $\s_3$
appears $N$ times along the diagonal.
The constraints on the pion field are now, in this basis,
\begin{eqnarray}
\label{constraintsgen}
\p_{NN}&=&-\sum_{i=1}^{N-1}\p_{ii}\ ,\\
\p_{ij}&=&(-1)^{i+j}\p_{ji}\ .\nonumber
\end{eqnarray}
In addition, $\p_{ii}$ is real for all $i$,
and $\p_{ij}=\p_{ji}^*$ for all $i\ne j$.
A minus sign in the pion propagator $\svev{\p_{ij}(x)\p_{ij}(y)}$,
\seef Eq.~(\ref{propexample}), occurs when $i$ is even and $j$ is odd,
or the other way around, because $J_{ii}J_{jj}=-1$ only in this situation.
Since minus signs in a field redefinition like Eq.~(\ref{fieldredef})
do not affect our arguments, we can choose
\begin{equation}
\label{fieldredefgen}
\hp_{ij}=i^{i+j}\p_{ij}\ .
\end{equation}
Now let us consider a diagram with $p_{ij}$ $\hp_{ij}$ propagators,
$n_{ij}$ $\hp_{ij}$ external lines,
and $v_{k,ij}$ vertices with $k_{ij}$ $\hp_{ij}$ lines attached to it.
Note that because of Eq.~(\ref{constraintsgen}) we can always take $i\le j$
(and $i\ne N$ if $i=j$, but this is not important).  We have that
\begin{equation}
\label{combgen}
2p_{ij}=n_{ij}+\sum_{k_{ij}} k_{ij}v_{k,ij}\ .
\end{equation}
This relation implies that a correlation function with $n_{ij}$ external
$\hp_{ij}$ lines equals $i^{-(i+j)n_{ij}}$ times the correlation function in terms of the
unhatted meson field $\p_{ij}$, using that $i^{-2p_{ij}}=i^{2p_{ij}}$, and Eq.~(\ref{combgen}).
The full correlation function changes by the product
\begin{equation}
\label{factor}
\prod_{ij}i^{-(i+j)n_{ij}}=i^{-\sum_{ij}(i+j)n_{ij}}\ ,
\end{equation}
where the product and sum are over all pairs $ij$ present in the correlation function.
The sum in the exponent on the right-hand side of Eq.~(\ref{factor}) always has to be
even, because every index has to appear an even number of times in the
correlation function for it not to vanish.   This means we can drop the minus sign
in this exponent, and we thus find the desired result.

Note that, unlike in the $N=1$ example, we do not always have that $n_{ij}$ is even.    A simple counter example
is the correlation function $\svev{\p_{12}\p_{23}\p_{34}\p_{41}}$, which does not
vanish, but has $n_{12}=n_{23}=n_{34}=n_{14}=1$.   However, clearly,
$(1+2)n_{12}+(2+3)n_{23}+(3+4)n_{34}+(1+4)n_{14}=20$ is even.

A similar type of argument was used in Ref.~\cite{GSS} to show the equivalence of
``standard'' quenched chiral perturbation theory \cite{BG} with ``non-perturbatively correct'' quenched chiral perturbation theory.

\section{\label{conc} Conclusion}
In QCD-like theories it is well known that physical observables
depend only on the effective vacuum angle $\theff$,
which is the difference between the explicit angle $\th$ multiplying
the topological term in the gauge-field action, and the (properly normalized)
$\Uone_A$ angle $\a$ of the fermion mass matrix.

When $N$ Dirac fermions belong to a real \irrep\ of the gauge group,
the theory can be reformulated in terms of $2N$ Majorana fermions.
The integration over a Majorana field yields a functional pfaffian.
As we discussed in the introduction, the phase of this pfaffian appears
to lead to a paradox: in certain cases, $\theff$ changes by
an odd multiple of $\p$ relative to its value in the initial Dirac theory.
Tracing the origin of this phenomenon we showed that, in the continuum,
the phase of the functional pfaffian is in fact inherently ambiguous,
as it depends on the choice of basis for the Majorana field.
A partial solution is that, in the case of $2N$ Majorana fermions,
one can fix the ambiguity
by appealing to the corresponding theory of $N$ Dirac fermions
in such a way that the apparent paradox is avoided.

A non-perturbative lattice definition of Majorana fermions
is free of the phase ambiguity: on any finite-volume lattice,
the (real-\irrep) Dirac operator becomes a finite-size matrix,
and, moreover, the lattice automatically selects the coordinate basis
to define the Dirac operator, and, hence, its pfaffian.
We reviewed the work of Ref.~\cite{SSt} who argued long ago that,
if the Wilson term in the Wilson lattice action for Dirac fermions is rotated
by a phase, that phase induces a topological term in the continuum limit.
We observed that there is a subtlety with this argument associated
with renormalization, which leads to a conjecture
(first made in Ref.~\cite{JS}) on how to complete the
argument of Ref.~\cite{SSt}, described in App.~\ref{SStconj}.
We generalized this result to domain-wall fermions, where this
subtlety does not arise,
as well as to the case of Majorana fermions.
This allowed us to unambiguously determine the effective vacuum angle,
finding consistent results between the Dirac and Majorana formulations
in all cases.

As an application, we discussed how chiral perturbation theory
reproduces the correct dependence on the explicit ($\th$) and
effective ($\theff$) vacuum angles.  This behavior has been long known
(even if maybe not widely known) for the effective theory for a gauge
theory with Dirac fermions,
but, to our knowledge, this is the first detailed study of this issue
for the effective theory for a gauge theory with Majorana fermions.
As such, our results fill in a lacuna in the discussion of Ref.~\cite{BL},
and resolve a question that was left open in Ref.~\cite{tworeps}.
In particular, we considered the chiral expansion for
$2N$ Majorana fermions in two cases that share $\theff=0$,
while the mass matrix is proportional to $J_S$ in one case,
and to $\id_{2N}$ in the other,
giving a diagrammatic proof that all physical observables
are equal in the two cases, as required by the common value of $\theff$.

\vspace{3ex}
\noindent {\bf Acknowledgments}
\vspace{2ex}

\noindent
We like to thank Steve Sharpe for useful discussions.
We also like to thank Jan Smit for comments and discussion on the
first version of this paper, which led to the addition of two new appendices.
The  work of MG is supported by the U.S. Department of
Energy, Office of Science, Office of High Energy Physics, under Award
Number DE-FG03-92ER40711.
YS is supported by the Israel Science Foundation
under grant no.~491/17.

\appendix
\section{\label{specrep} Spectral representation of the Dirac operator}
Consider the one-flavor Dirac operator for a general complex mass,
\begin{equation}
\label{Dgenm}
  D=\Sl{D}+me^{i\g_5\ta} \ ,
\end{equation}
where
\begin{equation}
\label{Dslash}
  \Sl{D} = -\Sl{D}^\dagger =
  \left(\begin{array}{cc} 0 & \bs_\m D_\m \\ \s_\m D_\m & 0
  \end{array}\right) \ ,
\end{equation}
with $\s_\m=(\id_2,i{\vec\s})$ and $\bs_\m=(\id_2,-i{\vec\s})$.
Let us derive the spectral representation of $\det(D)$
(see, for example, Ref.~\cite{LS}).
For a zero mode, depending on its chirality,
the eigenvalue is simply $me^{\pm i\ta}\equiv m_1 \pm i m_2$.
Turning to the non-zero modes we start with the right-handed spectrum
of the second-order operator,
\begin{equation}
\label{DsqR}
  -\Sl{D}^2 \j_R = - (\bs_\m D_\m) (\s_\n D_\n) \j_R = \l^2 \j_R \ ,
\end{equation}
where we take $\l$ real positive.  We consider the following {\it ansatz}
for an eigenvector of $D$:
\begin{equation}
\label{anzt}
  \left(\begin{array}{cc}
    me^{i\ta} & \bs_\m D_\m \\
    \s_\m D_\m & me^{-i\ta} \end{array}\right)
  \left(\begin{array}{c} \j_R \\ c \s_\n D_\n \j_R \end{array}\right)
  = \left(\begin{array}{c} (-c\l^2+me^{i\ta})\j_R \\
    (1+cme^{-i\ta}) \s_\m D_\m \j_R \end{array}\right) \ ,
\end{equation}
where the components of each column vector correspond to the two chiralities.
Requiring that the column vector on the left-hand side is an eigenvector gives rise to
a quadratic equation for $c$, with the two solutions
\begin{equation}
\label{solvealpha}
  c_\pm = \frac{im_2}{\l^2} \pm \frac{i}{\l} \sqrt{1+\frac{m_2^2}{\l^2}} \ .
\end{equation}
We denote the resulting eigenvectors by $\c_\pm$.
The corresponding eigenvalues are
\begin{equation}
\label{Dev}
  \l_\pm = m_1 \mp i\sqrt{\l^2+m_2^2} \ .
\end{equation}
The product of the two eigenvalues is $\l_+\l_-=\l^2+m^2$.
Remembering that there are $T$ zero modes per instanton, it follows that
the determinant of the one-flavor Dirac operator~(\ref{Dgenm}) is
\begin{equation}
\label{specDirac}
  \det(\Sl{D}+me^{i\ta\g_5})
  = (me^{i\ta})^{TQ} \prod_{\l>0} (\l^2+m^2) \ .
\end{equation}
The first factor on the right-hand side is the contribution
of the zero modes, where $Q$ is the topological charge of the
(multi-)instanton background field.  The second factor gives
the contribution of the non-zero modes in terms of the
eigenvalues of the second-order operator.
For the $N$-flavor case, substituting $\ta=\a/(NT)$ 
gives rise to Eq.~(\ref{DetF}).

\section{\label{bases} Majorana bases}
Let us consider for definiteness the case of a single Dirac fermion.
If we follow the basis transformations that lead to Eq.~(\ref{Dmaj3})
and then to Eq.~(\ref{finalWmaj}) we arrive at the following relations
\begin{eqnarray}
\label{JJ'}
  \J_1 &=& (\J'_1 - i\g_5\J'_2)/\sqrt{2} \ ,
\\
  \J_2 &=& (\J'_1 + i\g_5\J'_2)/\sqrt{2} \ ,
\nonumber
\end{eqnarray}
where the new Majorana fields $\J'_{1,2}$ correspond to the diagonal form of
the Dirac operator in Eq.~(\ref{finalWmaj}).
The original Dirac field can be expressed as
\begin{eqnarray}
\label{jJ'}
  \j &=& P_L \J_1 + P_R \J_2 \ = \ (\J'_1 + i\J'_2)/\sqrt{2} \ ,
\\
  \bj &=& (\bJ'_1 - i\bJ'_2)/\sqrt{2} \ .
\nonumber
\end{eqnarray}
where we have used Eq.~(\ref{JJ'}).  Equation~(\ref{jJ'}) is reminiscent
of the original notion of Majorana fields in Minkowski space as the real
and imaginary parts of a Dirac field, for a real \irrep.
As we explained in Sec.~\ref{majrev}, in this paper we prefer
the basis~(\ref{Diracreal}), because it respects the natural correspondence
between Weyl and Majorana fields.

We stress that the existence of the above basis, in which
the Majorana mass matrix is diagonal (see Eq.~(\ref{finalWmaj})),
does not resolve the puzzle we discussed in the introduction;
that puzzle must find its resolution when the basis~(\ref{Diracreal}) is used,
and indeed it does, as we showed in Sec.~\ref{latt}.

\section{\label{pospf} Positivity of the domain-wall pfaffian for positive mass}
In this appendix we prove that
the pfaffian of a single domain-wall Majorana fermion
is positive for a positive bare mass, \ie, in the absence of chiral phases.

We first consider the Dirac case.  We write the fermion action
as $S = \bj D_{DW}(m_0) \j$, where the explicit form of $D_{DW}(m_0)$
may be read off from Eq.~(\ref{Sdwf}).  It was proved in Ref.~\cite{YSdwf2}
that the partition function of a domain-wall fermion, $\det(D_{DW})$,
is strictly positive when $m_0>0$.
Turning to the case with one Majorana fermion, and introducing
the antisymmetric $\ca_{DW}(m_0)=R_5CSD_{DW}(U)$,
our task is to prove that $\pf(\ca_{DW}(m_0))$ is strictly positive as well.

The argument uses elementary calculus.
We consider a finite lattice, so that configuration space is compact.
Since $\det(D_{DW}(U))>0$ for any gauge-field configuration $U$,
it follows that there exists $\m>0$ such that $\det(D_{DW}(U)) \ge \m^2$,
for all $U$.
If this were not true, we could find an infinite sequence of configurations
$U_i$, such that $\lim \det(D_{DW}(U_i))\to 0$.  Because of compactness,
that sequence would have a convergent subsequence $U'_i \to U_{\rm lim}$,
where $U_{\rm lim}$ is a gauge configuration too.  It would then follow
that $\det(D_{DW}(U_{\rm lim}))=0$, contrary to the result that $\det(D_{DW}(U))>0$
for {\em all} configurations.

Next, we have $\pf^2(\ca_{DW}(U)) = \det(D_{DW}(U))$.
Therefore, either $\pf(\ca_{DW}(U)) > \m$ or $\pf(\ca_{DW}(U)) < -\m$.
Moreover, $\pf(\ca_{DW}(\id))$ is positive for the free case $U=\id$.

It follows that $\pf(\ca_{DW}(U)) > \m$ for all configurations $U$.
If this were not true, there would be a configuration $U_0$ for which
$\pf(\ca_{DW}(U_0))$ is negative.  Now choose a smooth path $U(t)$
such that $U(0)=\id$ and $U(1)=U_0$.  Along this path, $\pf(\ca_{DW}(U(t))$
must change continuously from positive to negative, and thus go through zero.
But, this is impossible, because we have seen that $|\pf(\ca_{DW}(U))|>\m$.

\section{\label{SStconj} Discussion of the result of Ref.~\cite{SSt}}
As was shown long ago in Ref.~\cite{KS} in the theory with $\th_W=\th_m=0$, the bare mass $m_0$ in Eq.~(\ref{Wth}) renormalizes additively.  This is related to the fact that there is no
symmetry distinguishing between the Wilson term $W$ and the single-site term proportional to $m_0$.
This fact was not considered in Ref.~\cite{SSt}.    The proof of the observation quoted in Eq.~(\ref{Zlim})
was given for a theory with Wilson fermions in the semi-classical limit, \ie, in the presence of a smooth background gauge field.   The
additive renormalization thus does not arise, as it is caused by quantum fluctuations of the gauge
field. In this appendix we describe a conjecture on the interplay of the
observation of Ref.~\cite{SSt} and renormalization.  Our discussion here
largely overlaps with Ref.~\cite{JS}.

With quantum effects thus ``mixing'' the Wilson and single-site mass terms in Eq.~(\ref{Wth}), the question
arises whether two angles $\th_W$ and $\th_m$ can be unambiguously introduced, and, if so, how this
should be done.   Here, we will discuss the issue, and formulate a natural conjecture answering this question.   A rigorous proof of our conjecture
is outside the scope of this paper.

First, consider a lattice gauge theory with Wilson fermions without any $\th$ angles
which leads to massless fermions in the continuum limit.
It follows from Ref.~\cite{KS} that in order to construct such a theory, the bare mass $m_0$ needs to be tuned to a critical value $m_c$ that depends on the bare coupling, \ie, the massless theory is obtained from a lattice theory with
fermion operator $D_K+W+m_c$.\footnote{The precise definition of $m_c$ is subject to ambiguities
of order $a^2$ \cite{Aoki}.   However, we will assume that we are in the scaling region, where these
ambiguities can be ignored.}
Using an axial rotation of the form~(\ref{rot5}), we can introduce an angle $\th_W$ in this theory,
turning the fermion operator into $D_K+e^{i\g_5\th_W}(W+m_c)$.  Of course, in the massless theory,
this angle has no physical consequence, consistent with what one expects in a massless continuum theory.   This construction does imply that if one starts with a theory with Wilson term
$e^{i\g_5\th_W}W$, the critical mass $m_c$ introduced to obtain a massless continuum limit also
needs to be multiplied by $e^{i\g_5\th_W}$.

Next, we may introduce a physical mass, which here we will take to be the axial-Ward-identity
(AWI) mass $m_{\rm AWI}$, by choosing
\begin{equation}
\label{AWI}
m_0=m_c+Zm_{\rm AWI}\ ,
\end{equation}
where we also introduced the multiplicative renormalization constant $Z$ relating the bare
subtracted lattice mass $m_0-m_c$ and the renormalized mass $m_{\rm AWI}$ \cite{KS,Betal}.
We may now introduce another angle $\th_m$ by considering the operator
\begin{equation}
\label{Diracren}
D_W(\th_W,\th_m)=D_K+e^{i\g_5\th_W}(W+m_c)+e^{i\g_5\th_m}Zm_{\rm AWI}\ .
\end{equation}
We recall that $m_c$ has already been determined for $m_{\rm AWI}=0$,
and that it is independent of both $\th_W$ and $\th_m$.  The question arises
how the $Z$ factor depends on these angles.
In order to address this question, we first apply an axial rotation~(\ref{rot5})
with $\eta=-\th_W/2$ to remove the phase of the Wilson term, arriving at
\begin{equation}
\label{Diracrenrot}
D'_W(\th_W,\th_m)=D_K+(W+m_c)+e^{i\g_5(\th_m-\th_W)}Zm_{\rm AWI}\ .
\end{equation}
At tree level, the Wilson-Dirac operator $D'_W$ is now function
of the difference $\th_m-\th_W$, and, by necessity, the same applies
to the $Z$ factor, order by order in perturbation theory.
The last step is to undo the axial rotation,
going from $D'_W$ back to $D_W$.  Assuming that our renormalization condition
transforms covariantly under axial rotations,\footnote{
  This includes as a special case any renormalization condition
  which is invariant under axial rotations of the fields.
}
the same $Z$ factor that we have determined for $D'_W$ will continue to satisfy
the corresponding renormalization condition for $D_W$.
It follows that, in Eq.~(\ref{Diracren}), and for general values
of $\th_W$ and $\th_m$, $Z$ is a function of the difference $\th_m-\th_W$ only.
We comment that the universal, logarithmic part of the $Z$ factor
is actually independent of $\th_W$ and $\th_m$.  However,
this $Z$ factor also has a finite part, and that part will in general
depend on $\th_W$ and $\th_m$, but, as we have just argued,
only through their difference.

Our conjecture is that in the fully dynamical theory
Eq.~(\ref{Zlim}) holds, with the fermion operator as defined in Eq.~(\ref{Diracren}).   We observe
that this conjecture is natural, in the sense that, in the continuum limit, the mass
$m_{\rm AWI}$ is the fermion mass $m$ to be used in Secs.~\ref{dirac}
and \ref{vacreal}.

\begin{boldmath}
\section{\label{theffchpt} $\th$ dependence of the chiral theory}
\end{boldmath}
As in Sec.~\ref{dirachpt} we consider here a gauge theory with $N$ Dirac fermions
in some complex \irrep.  For $\th=0$, the chiral lagrangian is constructed
using the non-linear field $U(x)\in SU(N)$, see Eq.~(\ref{Upion}).
We will prove that, at both leading order (LO) and next-to-leading order (NLO), the chiral lagrangian for $\th\ne 0$
is obtained via the replacement $U(x)\to \S(x)$, where $\S(x)=U(x)\S_0$
(see Eq.~(\ref{Upion})), and where $\S_0$ is given by Eq.~(\ref{Sig0}).\footnote{%
  We conjecture that a similar statement applies to all orders
  in the chiral expansion.
}
As before, $\cm$ is given in Eq.~(\ref{cm}).

We start at tree level.  The requirement that the lagrangian
of the chiral theory depend on $\th$ and $\a$ only through
their difference $\theff$ is satisfied if the potential admits the form
\begin{equation}
\label{Vclf}
  V = - \frac{f^2 B}{2} \tr(e^{if(\theff)}\cm^\dagger \S + \hc) \ .
\end{equation}
This amounts to
multiplying $\cm^\dagger \S$ in Eq.~(\ref{Vcl})
by the phase factor $e^{if(\theff)}$, where $f(\theff)$ is {\em a-priori}
an arbitrary (real) function of its argument.

We first invoke the chiral power counting, which implies
that the tree-level lagrangian should be linear in $\cm$ or $\cm^\dagger$.
This dependence is already explicit in Eq.~(\ref{Vclf}), and so
\begin{equation}
\label{ftheff}
  f(\theff) = f(\th-\a) = f(\th-\Im\log\det\,\cm) \ ,
\end{equation}
must in fact be independent of $\cm$.  This allows us to set $f(\theff)=c$
in Eq.~(\ref{Vclf}), where $c$ is some constant.

Next we consider the special case where $\cm=m\id_N$, with $m>0$, and $\th=0$.
Now $\S=U$ and the tree-level lagrangian must be invariant under
the (internal) parity transformation $U(x)\to U^\dagger(x)$.
This invariance is respected only for $e^{ic}=\pm 1$,
which completes the argument.\footnote{
  The choice $e^{ic}=+1$ is conventional.
}

The reasoning at NLO is similar.
{\em A-priori}, $\cm^\dagger \S$ can again be multiplied by a phase factor
$e^{if(\theff)}$, with $f(\theff)$ a new arbitrary real function
for each occurrence of $\cm^\dagger \S$.  But, as before, the power counting
restricts every such $f(\theff)$ to a constant.
In the last step we consider the most general constant phase factors
consistent with parity invariance of the $\cm=m\id_N$, $\th=0$ theory,
finding that this does not give rise to any new operators not already
present in the standard NLO chiral lagrangian.

\vspace{3ex}
\begin{boldmath}
\section{\label{proofSigLO} Proof of Eq.~(\ref{SigLO})}
\end{boldmath}
Let us prove, algebraically, that the global minimum
of the tree-level potential~(\ref{Vcl}) is given by $\S_L$ of Eq.~(\ref{SigLO}),
with $n=n(\theff)$ as described in Sec.~\ref{dirachpt}.

We begin by writing the $\SU(N)$ matrix $U$ of Eq.~(\ref{Upion}) as $U=\tO\tU$,
so that the potential becomes
\begin{equation}
  V = -\frac{f^2 B m}{2} \tr(e^{i\theff/(NT)}\tU + \hc) \ .
\label{VclnotO}
\end{equation}
We may assume without loss of generality that $\tU$ is diagonal,
\begin{equation}
\label{tU}
  \tU = diag(e^{i\f_1},e^{i\f_2},\ldots,e^{i\f_{N-1}},e^{i\f_N})\ ,
\end{equation}
where $\f_1,\ldots,\f_{N-1}$ are the independent real variables, and
\begin{equation}
\label{phiN}
  \f_N = 2\p n - (\f_1+\ldots+\f_{N-1}) \ ,
\end{equation}
with $n$ an arbitrary integer. Introducing the shorthand $\tilt=\theff/(NT)$
we need to find the global maximum of
\begin{equation}
\label{Vflipped}
  \cv = \half \tr(e^{i\tilt}\tU + \hc) = \sum_{k=1}^N \cos(\tilt+\f_k) \ .
\end{equation}
The saddle-point conditions are
\begin{equation}
\label{Vsaddle}
  \sin(\tilt+\f_k) = \sin(\tilt+\f_N) \ , \qquad k=1,2,\ldots,N-1 \ .
\end{equation}
First consider a solution with all phases equal.
Equation~(\ref{phiN}) then implies that $\f_k=2\p n/N$, $k\in\{1,\dots,N\}$,
for some $n$, and $\cv=N\cos(\tilt+2\p n/N)$.
The global maximum over this set of solutions is obtained
for $n=n(\theff)$, defined as before as the value of $n$ for which
$\tilt+2\p n/N$ is closest to zero.  The value of this maximum is
\begin{equation}
\label{cvmax}
  \cv_{\rm max} = N \cos(\theff/(NT)+2\p n(\theff)/N) \ ,
\end{equation}
which reproduces Eq.~(\ref{Vmin}).

It remains to prove that this solution is in fact the global
maximum of $\cv$ over the entire set of saddle points.
What complicates matters is that Eq.~(\ref{Vsaddle}) can be satisfied by
$\f_k=\f_N$, or by $\f_k=\p-2\tilt-\f_N$.  In the former case we have
$\cos(\tilt+\f_k) = \cos(\tilt+\f_N)$, whereas in the latter case
we have $\cos(\tilt+\f_k) = -\cos(\tilt+\f_N)$, so that
$\cos(\tilt+\f_k) + \cos(\tilt+\f_N) = 0$.

Let us denote by $\cv_{\rm max}^{(1)}$ the maximal value of $\cv$
when $\f_1=\p-2\tilt-\f_N$, while the remaining $N-2$ independent phases
are equal to $\f_N$.  It follows immediately from the above discussion that
in this case $\cv=\sum_{k=3}^N \cos(\tilt+\f_k)=(N-2)\cos(\tilt+\f_N)$,
leading to the upper bound
\begin{equation}
\label{cv1}
  \cv_{\rm max}^{(1)}\le N-2 \ .
\end{equation}
Similarly, if exactly two independent phases
are equal to $\p-2\tilt-\f_N$, then the corresponding maximal value
is bounded by $\cv_{\rm max}^{(2)}\le N-4$, and so on.

We also need a lower bound on the maximum in Eq.~(\ref{cvmax}).
The maximum value the angle $\tilt+2\p n(\theff)/N$ can take is equal to
$\p/N$.
Since $\sin(x)\le x$, this implies
\begin{equation}
\label{boundmax}
  \cos(\theff/(NT)+2\p n(\theff)/N) \ge \sqrt{1-(\p/N)^2} \ .
\end{equation}
It follows that $\cv_{\rm max}$
of Eq.~(\ref{cvmax}) is larger than $\cv_{\rm max}^{(1)}$ if
\begin{equation}
\label{boundN}
  N \sqrt{1-(\p/N)^2} \ge N-2 \ ,
\end{equation}
which is true for $N\ge 4$.

It remains to check explicitly the cases $N=2,3$.
For $N=2$, choosing $\f_1=\p-2\tilt-\f_2$ gives $\cv=0$,
which is smaller than the maximum in Eq.~(\ref{cvmax}).\footnote{An
exception is the case $\tilt=\p/2$, for which
Eq.~(\ref{cvmax}) vanishes, too.  In fact, the tree-level potential
is identically zero in this case \cite{Smilga}.  For $\tilt$ close to $\p/2$
there is competition between LO and NLO, and the above discussion
does not apply.}

For $N=3$, if we choose $\f_1=\p-2\tilt-\f_3$ and $\f_2=\f_3$
then $\cv_{\rm max}^{(1)}\le 1$ according to the upper bound~(\ref{cv1}).
By contrast, for the solution with $\f_1=\f_2=\f_3$,
the maximum value the angle $\tilt+2\p n(\theff)/N$ is now $\p/3$;
hence, $N\cos(\theff/(NT)+2\p n(\theff)/N)$ is bounded from
below by $3/2$, making $\cv_{\rm max}$ again the true global maximum.

\vspace{3ex}

\end{document}